\documentclass[granma]{svjour}
\usepackage{graphicx}
\usepackage{dcolumn}
\usepackage{bm}
\usepackage{epsfig}
\usepackage{psfrag}
\usepackage{color}
\usepackage{fancybox}
\usepackage{float}
\usepackage{ulem}
\newcommand{\dkc}[1]{#1}

\begin{document}

\title{Granular packings of cohesive elongated particles}

\author{R.C. Hidalgo \inst{1}\mail{raul.cruz@udg.edu}, D. Kadau \inst{2}, 
T. Kanzaki \inst{3}  and H.J. Herrmann \inst{2,4}}

\institute{ \inst{1} 
\inst{1} Departamento de F\'{\i}sica y Matem\'atica Aplicada, 
Universidad de Navarra,  31080 Pamplona , Spain. \\
\inst{2} Departament de F\'{\i}sica, Universitat de Girona, 17071 Girona, Spain \\
\inst{3}Institute for Building Materials, ETH Z\"urich,
8093 Z\"urich, Switzerland\\ 
\inst{3}Departamento de F\'{\i}sica,
Universidade Federal do Cear\'a, 60451-970 Fortaleza, Cear\'a,
Brazil }

\date{\today}

\maketitle
\begin{abstract}
We report numerical results of  effective attractive forces
 on the packing properties of two-dimensional elongated grains.  In
deposits of non-cohesive rods in 2D, the topology of the packing is
mainly dominated by the formation of ordered structures of aligned rods.
Elongated particles tend to align horizontally and the stress is mainly
transmitted from top to bottom, revealing an asymmetric distribution of
local stress. However, for deposits of cohesive particles, the preferred
horizontal orientation disappears.  Very elongated particles with strong
attractive forces form extremely loose structures, characterized by an
orientation distribution, which tends to a uniform behavior when
increasing the Bond number. As a result of these changes, the pressure
distribution in the deposits changes qualitatively.  
The isotropic part of the local stress is notably enhanced with respect to the 
deviatoric part, which is related to the gravity direction. Consequently, the lateral stress
transmission is dominated by the enhanced disorder and leads to a faster
pressure saturation with depth.

\keywords{Granular matter, Molecular Dynamics, non-spherical particles
  Quicksand, Collapsible soil}
\end{abstract}

\section{Introduction}
Nowadays, granular materials have remarkable relevance in engineering and physics 
\cite{review1,review2,review3}. During the last decades, 
important experimental and theoretical efforts have been made for better understanding 
the mechanical behavior of these many-body systems. Specifically, we
highlight the career of {\bf Professor Isaac Goldhirsch}, who 
has made a considerable number of remarkable contributions in this field \cite{isaac1,isaac2,isaac3,go1,go2}. 

\begin{figure}
 \begin{center}
\includegraphics[width=6.9cm]{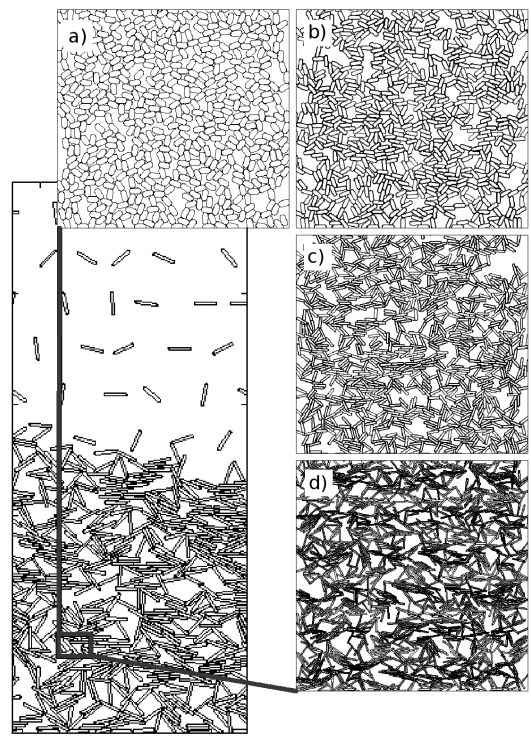}
 \caption[]{ Simulated packings of elongated cohesive particles settled
 by gravity. Final configurations are shown for the same granular bond number $Bo_g=10^4$ 
and increasing elongation: (a) $d=2$, (b) $d=3$, (c) $d=5$ and (d) $d=10$.}
 \label{fig:snap}
\end{center}
\end{figure}
Despite the fact that granular materials are often composed of 
particles with anisotropic shapes, like rice, lentils or pills, most of the experimental and
theoretical studies have focused on spherical
particles~\cite{review1,review2,review3}. 
In recent years several studies have highlighted the qualitatively 
new features induced by particle shape~\cite
{gallas93a,marroquin08a,marroquin09a,marroquin09b,villarruel00a,donev04a,zhong01a,blouwolff06a,desmond06a,sandpileani}. 
These include effects in the packing 
fraction of granular piles~\cite{villarruel00a,donev04a}, the pressure in
the lateral walls of a silo during its discharge~\cite{zhong01a},
the mean coordination number~\cite{blouwolff06a}, the
jamming~\cite{desmond06a} and the stress propagation in granular
piles~\cite{sandpileani}. Moreover, there is currently increasing
interest on the effect that particle's shape has on the global
behavior of granular materials ~\cite{lumay06a,azema07a,azema09a,azema10a,raulprl2009,raul10a}. 

Very often, loose  and disordered granular structures 
appear in many technological processes and even everyday life. 
They can be found in  collapsing soils \cite{Mitchell05,Barden1973,Assallay1997,Reznik05}, 
fine powders \cite{Rognon06,kadau2003a,roeck2008b} or 
complex fluids \cite{wagner2009,delgado2010}. Generally, those fragile 
grain networks are correlated with the presence of cohesive forces \cite{Rognon06,kadau2009a,kadau2009b,kadau2009c}.  
Typical fine powders have in most cases an effective attractive 
force, e.g. due to a capillary bridge between the particles or 
van der Waals forces  (important when going to very small grains, e.g.
nano-particles). This force is known to be of great importance, e.g.\ for the
mechanical behavior and the porosity of the macroscopic material
\cite{kadau2010a,kadau2003b,kadau2002,bartels2005}. This cohesive force leads
also to the formation of loose and fragile granular packings as investigated
in detail for structures generated by successive deposition of spherical grains
under 
the effect of gravity \cite{kadau2009a,kadau2009b,kadau2010a,kadau2011}. 

The main objective of this work is to clarify the effect that an 
effective attractive force has on the packing properties of elongated
grains.  Here we focus on packings generated by deposition under 
gravity. The paper is organized as follows: in Section \ref{model} we review
the theoretical model used in the numerical simulations. The results on the deposit structure as well as the details of the packings' 
micro-mechanics are presented and discussed in Section ~\ref{Results}. 
Finally, there is a summary with conclusions and perspectives.

\section{Model}
\label{model}
We have performed Discrete Element Modeling of a
two-dimensional granular system composed of \dkc{ non-deformable oval 
particles, i.e.\  spheropolygons \cite{marroquin08a,marroquin09a} composed of two lines of equal length and two half circles
of same diameter. The width of a particle is the smaller diameter, given by
the distance between the two lines (equals the circle diameter), whereas the length
is the maximum extension. The aspect ratio $d$ is defined by the length
divided by width.}   This system is confined within a rectangular 
box of width $W$. Its lateral boundaries as well as the bottom are each built
of one \dkc{ very long spheropolygonal}  particle, which is fixed.
In order to generate analogous deposits, the system width is always set to
$W=20 \times d$ (in units of particle width).
As illustrated in Fig.\ \ref{fig:snap} the particles are continuously added at the top of the box with very low feed rate and 
a random initial velocity and orientation. 
The granular system settles  under the effect of gravity and is
relaxed until the particles' mean kinetic energy is several
orders of magnitude smaller than its initial value. 

In the simulation, each particle $i$ $(i=1...N)$ has three degrees of freedom,
two for the translational motion and one for the rotational one. The
particles' motion is governed by Newton's equations of motion 
\begin{eqnarray}
m \ddot{\vec{r}}_i = \sum^{c}_j \vec{F}_{ij} - m g~\hat{\vec{e}}_y, 
~~~~ I \ddot{\theta}_i  = \sum^{c}_j (\vec{l}_{ij} \times
     \vec{F}_{ij})\cdot \hat{\vec{e}}_z,
\label{eq:newton}
\end{eqnarray}
\noindent where $m$ is the mass of  particle $i$, $I$ its momentum of inertia, 
$\vec{r}_{i}$ its position and $\theta_{i}$ its rotation angle.  
$g$ is the magnitude  of the  gravitational field and $\hat{\vec{e}}_y$ is 
the unit vector in  the vertical direction.  
$\vec{l}_{ij}$ is the vector from the center of mass of particle $i$ pointing
to the contact point, $\hat{\vec{e}}_z$ is the normal vector in $z$-direction
(perpendicular to the simulation plane).
In Eq.\ref{eq:newton}
$\vec{F}_{ij}$ accounts for the force exerted by particle $j$ on $i$ and 
it can be decomposed  as
$\vec{F}_{ij} = F_{ij}^{N}\cdot\hat{\vec{n}} + F_{ij}^{T}\cdot\hat{\vec{t}},$
\noindent where $F_{ij}^{N}$ is the component in normal direction
$\hat{\vec{n}}$ to the contact plane. Complementary, $F_{ij}^{T}$ is the component acting in tangential direction $\hat{\vec{t}}$. 
For calculating the particles' interaction $\vec{F}_{ij}$ we use a very efficient algorithm
proposed recently by Alonso-Marroqu\'{i}n et al \cite{marroquin08a,marroquin09a}, 
allowing for simulating a large number of particles. 
This numerical method is based on the concept of spheropolygons, where the 
interaction between two contacting particles only is governed by the overlap distance 
between them (see details in Ref.\ \cite{marroquin08a,marroquin09a}). 
To define the normal interaction $F_{ij}^{N}$, we use a nonlinear Hertzian 
elastic force \cite{landau86a}, proportional to the overlap distance $\delta$ of the particles. 
Moreover, to introduce dissipation, a velocity dependent viscous damping is
assumed.  Hence, the total normal force reads as
  $F_{ij}^{N}=-k^{N}\cdot\delta^{3/2}-\gamma^{N}\cdot v_{rel}^{N}$,
where $k^{N}$ is the spring constant in the normal direction, $\gamma^{N}$ is
the damping coefficient in the normal direction and  $v_{rel}^{N}$ is the normal relative velocity between $i$ and $j$. 
The tangential force $F_{ij}^{T}$ also contains 
an elastic term and a tangential frictional term accounting also for static 
friction between the grains.  
Taking into account Coulomb's friction law it reads as,
$F_{ij}^{T}=\min\{-k^{T}\cdot\xi-\gamma^{T}\cdot|v_{rel}^{T}|,\,\mu
      F_{ij}^{N} \}$,
where $\gamma^{T}$ is the damping coefficient in tangential direction,
$v_{rel}^{T}$  is the tangential component of the relative contact
velocity of the overlapping pair. $\xi$ represents the elastic elongation
of an imaginary spring with spring constant $k^T$ at the contact \cite{cundall79a}, which increases as
$d\xi(t)/dt = v_{rel}^{T}$ as long as there is an overlap between the
interacting particles \cite{cundall79a,duvaut72a}. $\mu$ is the friction
coefficient of the particles. 

Additionally, here we consider bonding between two particles in terms of a
cohesion model with a constant attractive force $F_c$ acting within a finite range $d_c$. 
Hence, it is expected that the density and the characteristics 
of the density profiles are determined by the ratio between the cohesive force $F_c$ and gravity $F_g=mg$ , 
typically defined as the granular Bond number $Bo_g = F_c/F_g$. Thus, the case of $Bo_g=0$ corresponds to the cohesion-less case whereas for $Bo_g \rightarrow \infty$  gravity is negligible. 

The equations of motion, Eqs.\ref{eq:newton}, are integrated using a fifth order predictor-corrector algorithm with a numerical error proportional to $(\Delta t)^{6}$ \cite{allen87a}, while the kinematic tangential displacement, 
is updated using an Euler's method. In order to model hard particles, the maximum overlap must always be
much smaller than the particle size; this is ensured by introducing 
values for normal and tangential elastic constants, $k^T/k^N=0.1$, $k^N=10^{3} N/m^{3/2}$.  
The ratio between normal and tangential  damping coefficients is taken as $\gamma^N/\gamma^T=3$,  
$\gamma^T=1\times10^2 s^{-1}$ while gravity is set to $g=10~m/s^2$ and the
  cohesion range to $d_c=0.0001$ (in units of particle width) to account for a
  very short range attraction, as mediated, e.g. by capillary bridges or
  van der Waals force. 
For these parameters, the time step should be around $\Delta t= 5 \times
10^{-6}s$.  In all the simulations reported here, we have kept 
the previous set of parameters and only the particle aspect ratio and the Bond
number $Bo_g$   
have been modified. We have also carried out additional runs (data not shown)
using other particles' parameters, and  we have verified that the trends  and
properties of the quantities we subsequently analyze are robust to such
changes.  

\section{Results and discussion}
\label{Results}
\begin{figure}
 \begin{center}
\includegraphics[width=4.3cm]{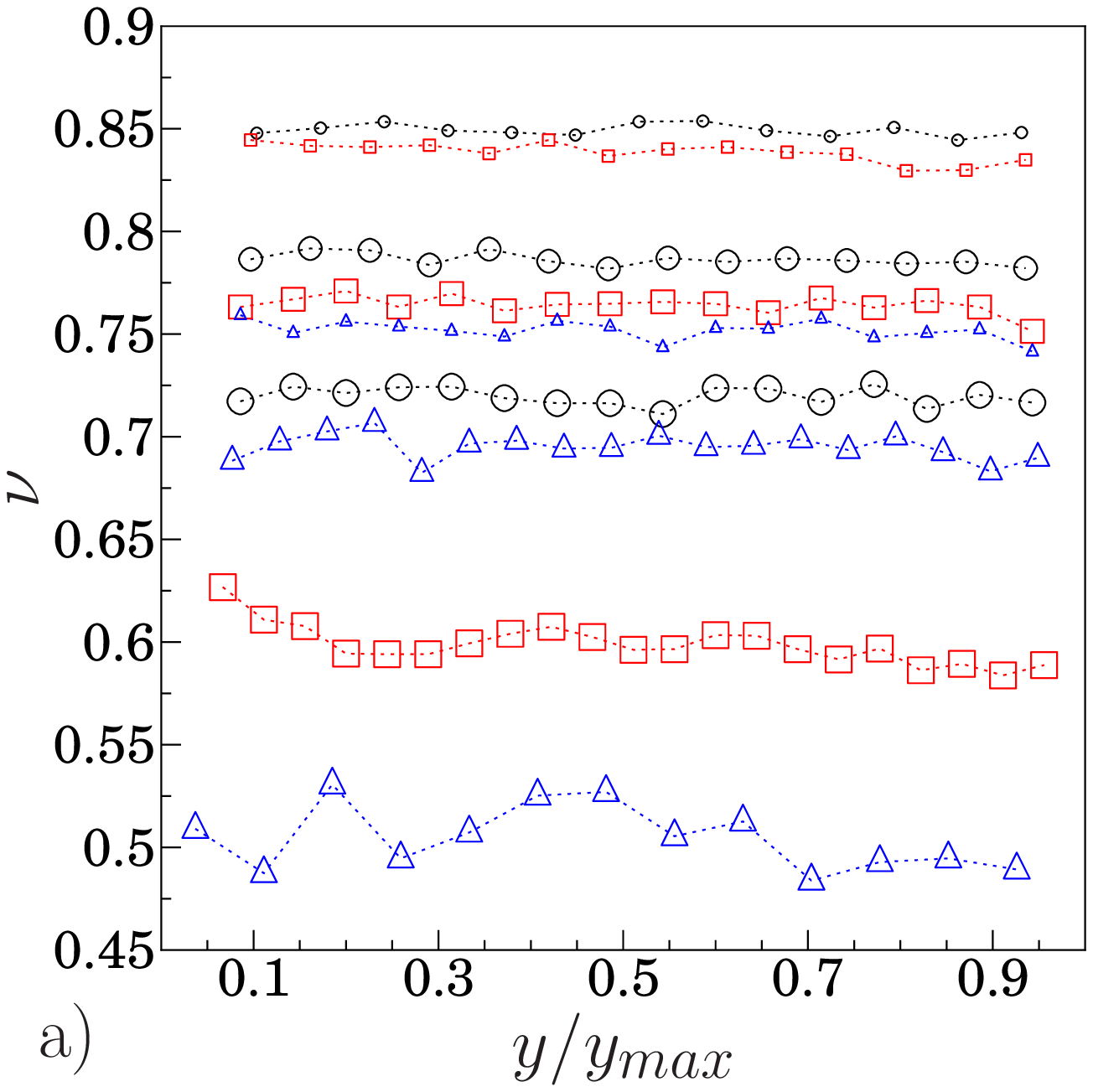}
~\includegraphics[width=4.3cm]{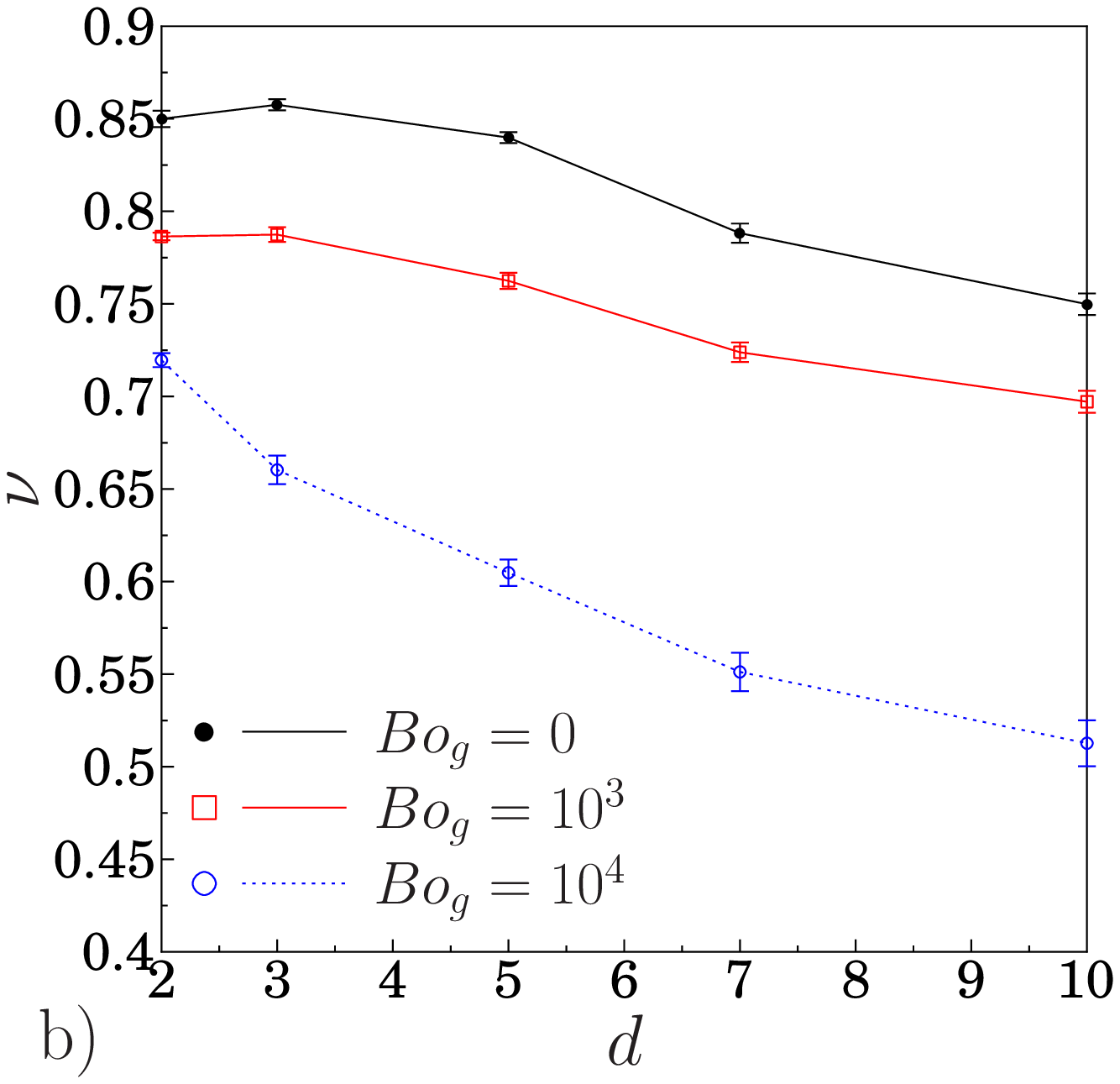}
   \caption{ (Color online) Density profiles of different granular deposits.
In a) the circles represent particles with $d=2$, the squares $d=5$ and the
triangles $d=10$. 
 The larger the symbols the stronger is the attractive force
  (small: $Bo_g=0$; medium: $Bo_g=10^3$; large: $Bo_g=10^4$).    
In b) the evolution of the average volume fraction as a function of $d$ is shown.}
\label{density_profiles}
\end{center}
\end{figure}
We systematically study granular deposits of particles with aspect 
ratio from $d=2$ to $d=10$  and different Bond numbers. In all simulations 
presented here we have used $6\times10^{3}$ rods. 
In Fig.~\ref{fig:snap} we illustrate
 the granular packings obtained for several particle shapes and constant 
Bond number $Bo_g=10^4$.   
Despite of the presence of a gravity field acting downwards,  
the formation of very loose and  disordered granular structures 
is very noticeable. 
Moreover, as the aspect ratio of the particles increases, 
the volume fraction of the column decreases, 
showing a tendency to the formation of more disordered structures.   
This result contrasts with what was obtained for non-cohesive 
elongated particles. In that case, the topology of the packing is dominated by
the 
face to face interaction and the formation of ordered structures of aligned
rods is detected \dkc{ \cite{berntsen,raulprl2009}.}  

For better describing the packing structure, in Fig\ref{density_profiles}a 
we present the density profiles depending on depth $y/y_{max}$ obtained for several 
deposits of elongated particles. We plot for each particle aspect 
ratio density profiles with increasing strength of the attractive force
illustrated in Fig\ref{density_profiles}a by increasing symbol size. 
In systems composed by elongated particles with strong 
attractive forces the formation of extremely loose structures is observed,
which are stabilized by the cohesive forces \cite{kadau2011}. 
 Moreover, in all cases the density profiles are quite 
 uniform as function of depth.  
Typically, for the non-cohesive case a close packing is expected. 
For cohesive particles, smaller volume fraction values are
found and the density profiles  remain constant with depth. These density profiles have been studied and analyzed extensively  for 
spherical particles \cite{kadau2011} where \dkc{  constant density has
  been found only for fast deposition as strongly influenced by
  inertia. However, an
  extremely  slow and gentle deposition process, allowing for 
relaxation of the deposit due to its own weight after each 
deposition, leads to a decreasing density with vertical position. Obviously,
here the feed rate is not sufficiently slow. Additionally, as particles are
added at the top they are accelerated before, thus reaching the deposit with a
non negligible impact velocity. } 
 
\begin{figure}
 \begin{center}
\includegraphics[width=4.3cm]{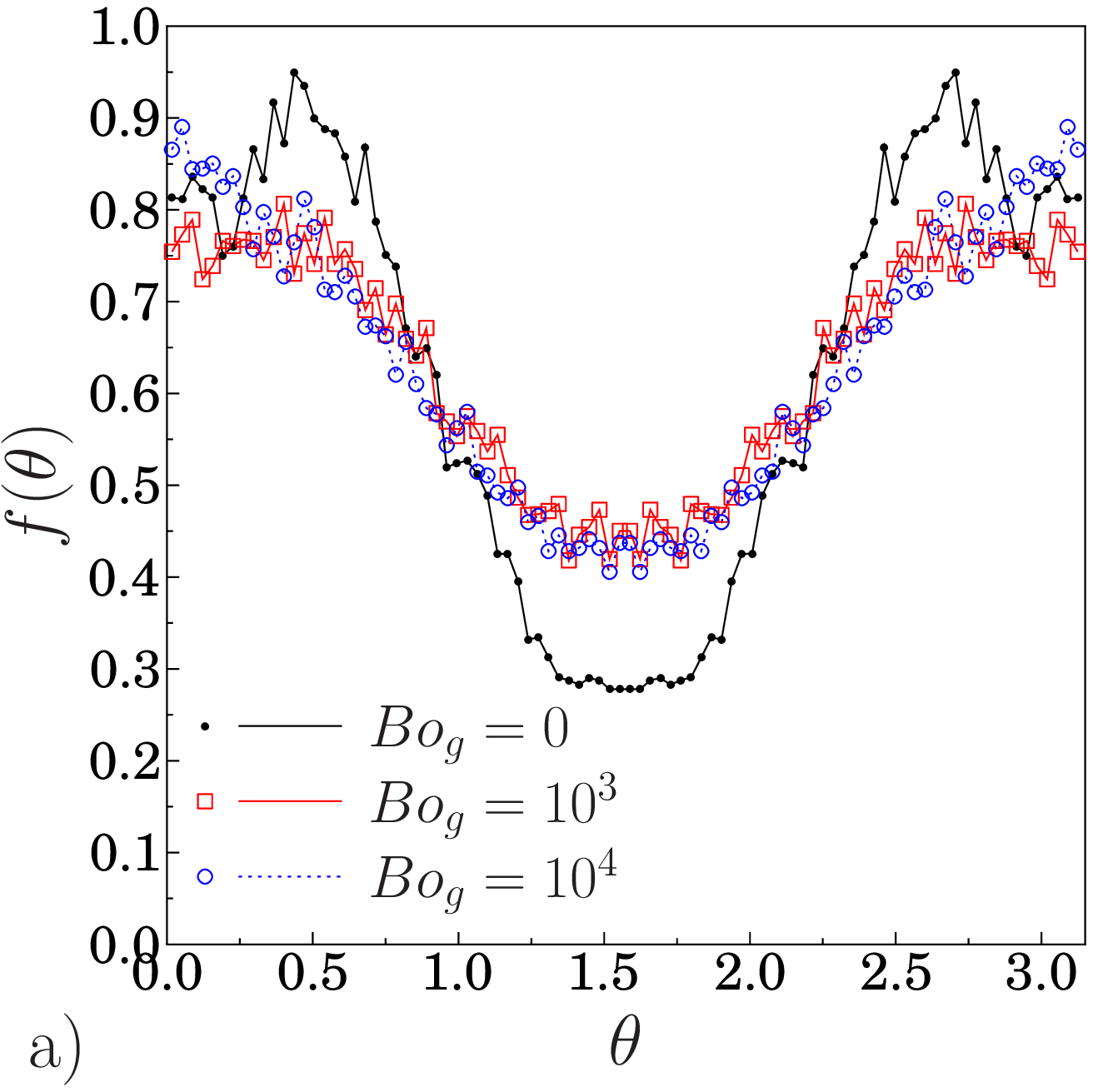}
~\includegraphics[width=4.3cm]{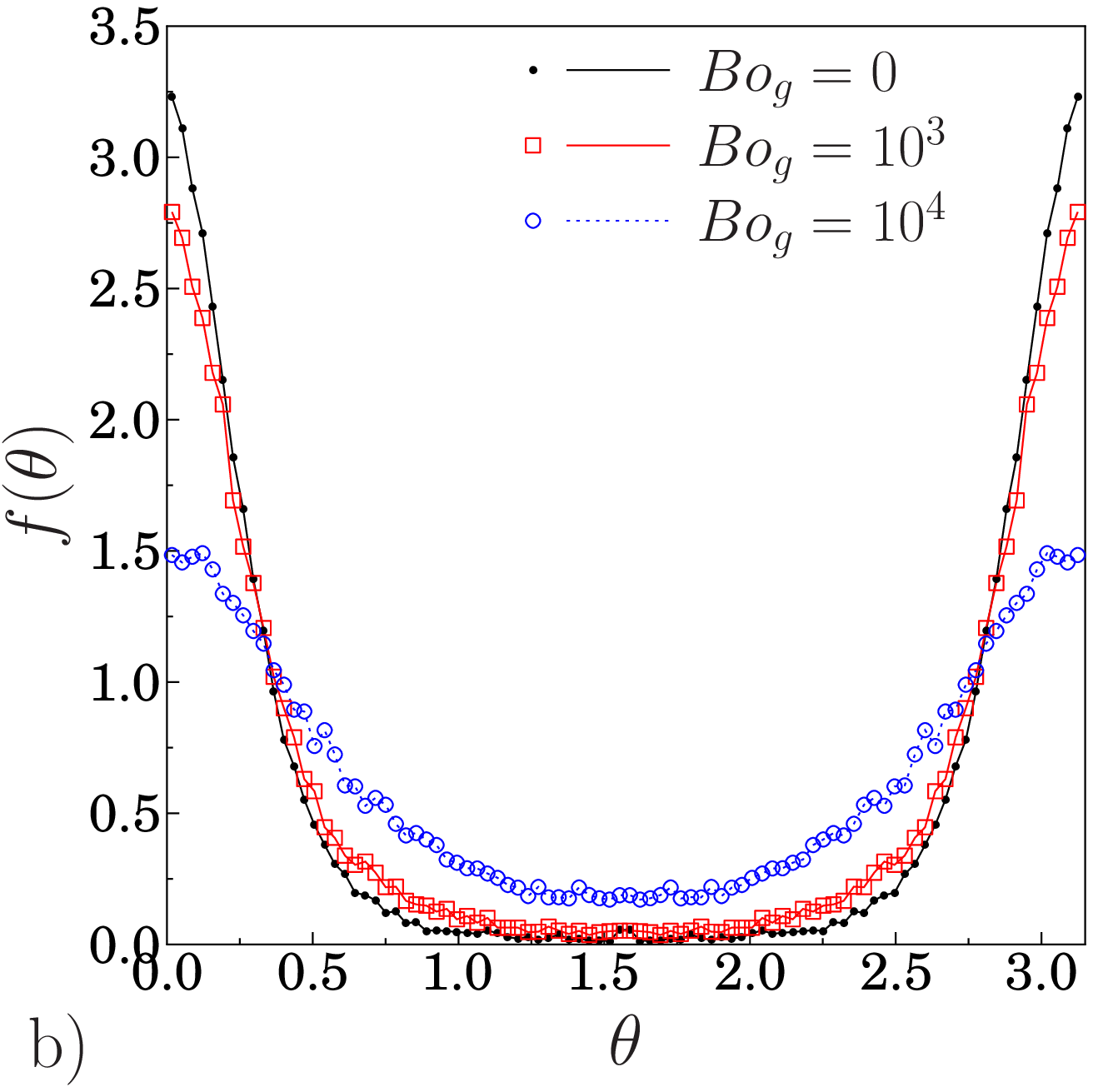}
   \caption{ Orientation distributions of particles for two  aspect ratios, a) $d=2$ and b) $d=10$. In each case, results for several bond numbers are presented.}
\label{fig:distribu_op} 
\end{center}
\end{figure}
Complementary, in Fig\ref{density_profiles}b a systematic study of 
the global volume fraction depending on the particle aspect ratio is presented
for different bond numbers. All curves  show the overall trend
of decreasing density with increasing aspect ratio $d$.
\dkc{  For packings of non-cohesive particles disordered and thus substantially looser structures can only be found with
very large aspect ratio as found earlier numerically and experimentally \cite{raulprl2009,raul10a}.}
For very cohesive particles (see Fig.\ref{fig:snap}), however, loose and disordered granular 
structures can be  easily stabilized, leading to much lower densities
independent on shape/elongation.  This is notably enhanced 
as the aspect ratio of the particles increases and, consequently, the volume 
fraction of the packings quickly decreases.

In order to characterize the packing morphology, we 
examine the orientations of the particles. 
In figure \ref{fig:distribu_op}, the distributions of particle's orientation
$f(\theta)$, with respect to the horizontal direction, 
for rods with aspect ratio $d=2$ and $d=10$ are illustrated. 
We present results for several bond numbers. First, 
for the non-cohesive  case, the geometry of the particle 
dominates the final structures of the compacted piles. 
Note that at the end of the deposition process, long particles ($d=10$) 
most probably lie parallel to the substrate ($\theta=0$ and $\theta=\pi$), while the most
unlikely position corresponds to standing rods ($\theta=\frac{\pi}{2}$). 
Nevertheless, as the aspect ratio decreases, there is a shift in the most 
probable orientation, leading to a peaked distribution at an
  intermediate orientation \cite{raulprl2009,raul10a}.
 \dkc{ Furthermore, this
  shift of the maximum is not observed for highly cohesive particles.
In general, as the strength of the attractive force gets stronger, 
the final packing tends to a flatter distribution. }
As we pointed out earlier, 
very elongated particles with strong attractive forces form extremely 
loose structures. As the aspect ratio gets higher, the rods form highly 
jammed and disordered networks 
 because during the deposition local particle 
rearrangements are constrained by the attractive force. 
The latter, is corroborated by the distribution of the angular orientation,
where now the probability for standing rods which was zero in the non-cohesive
case is enhanced with increasing bond number
(fig.\ \ref{fig:distribu_op}b). It seems that, for sufficiently large aspect
ratio $d$,  there is a threshold bond number
needed to have a non-zero probability for standing rods.

\begin{figure}
 \begin{center}
\includegraphics[width=4.3cm]{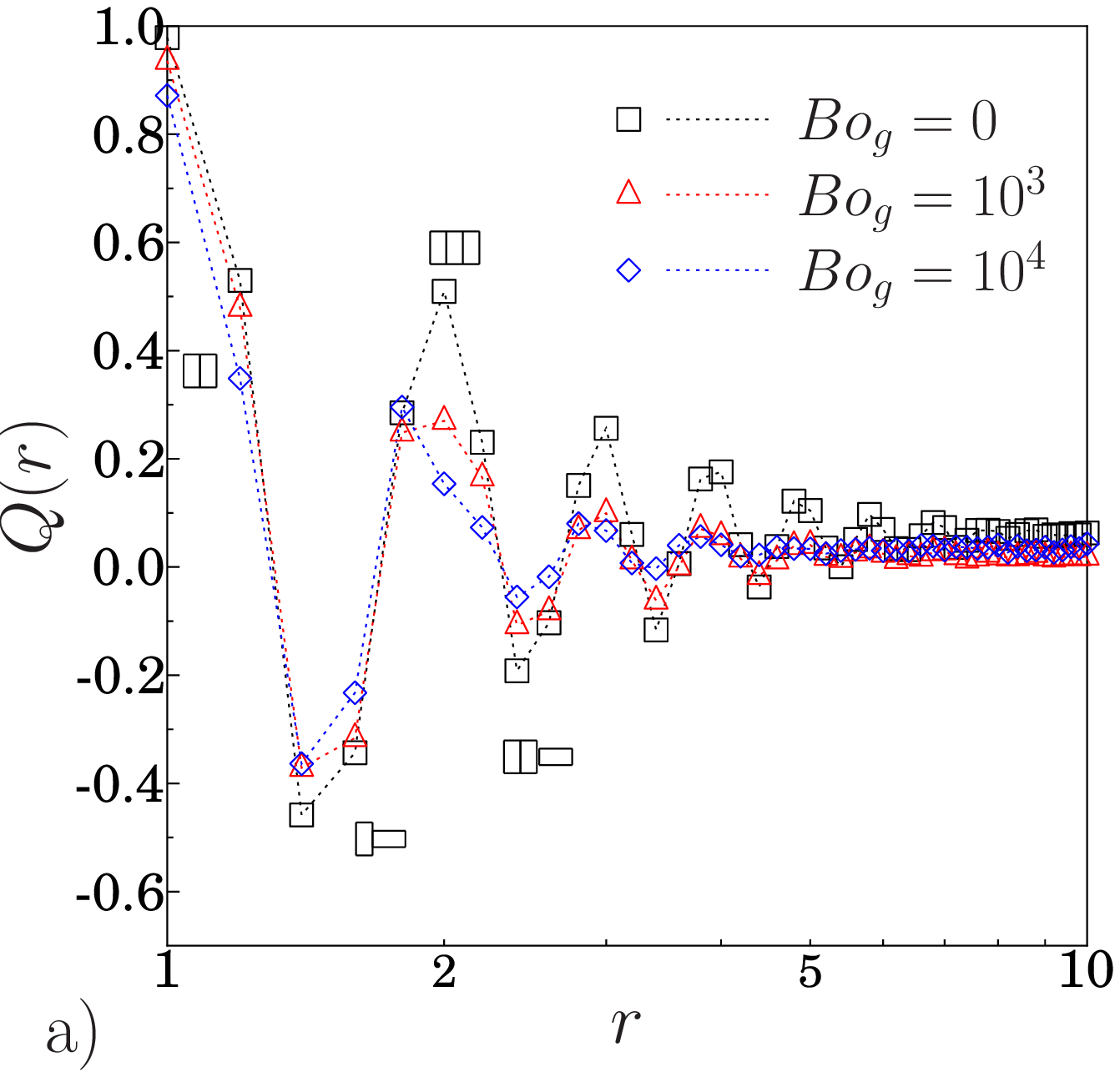}
~\includegraphics[width=4.3cm]{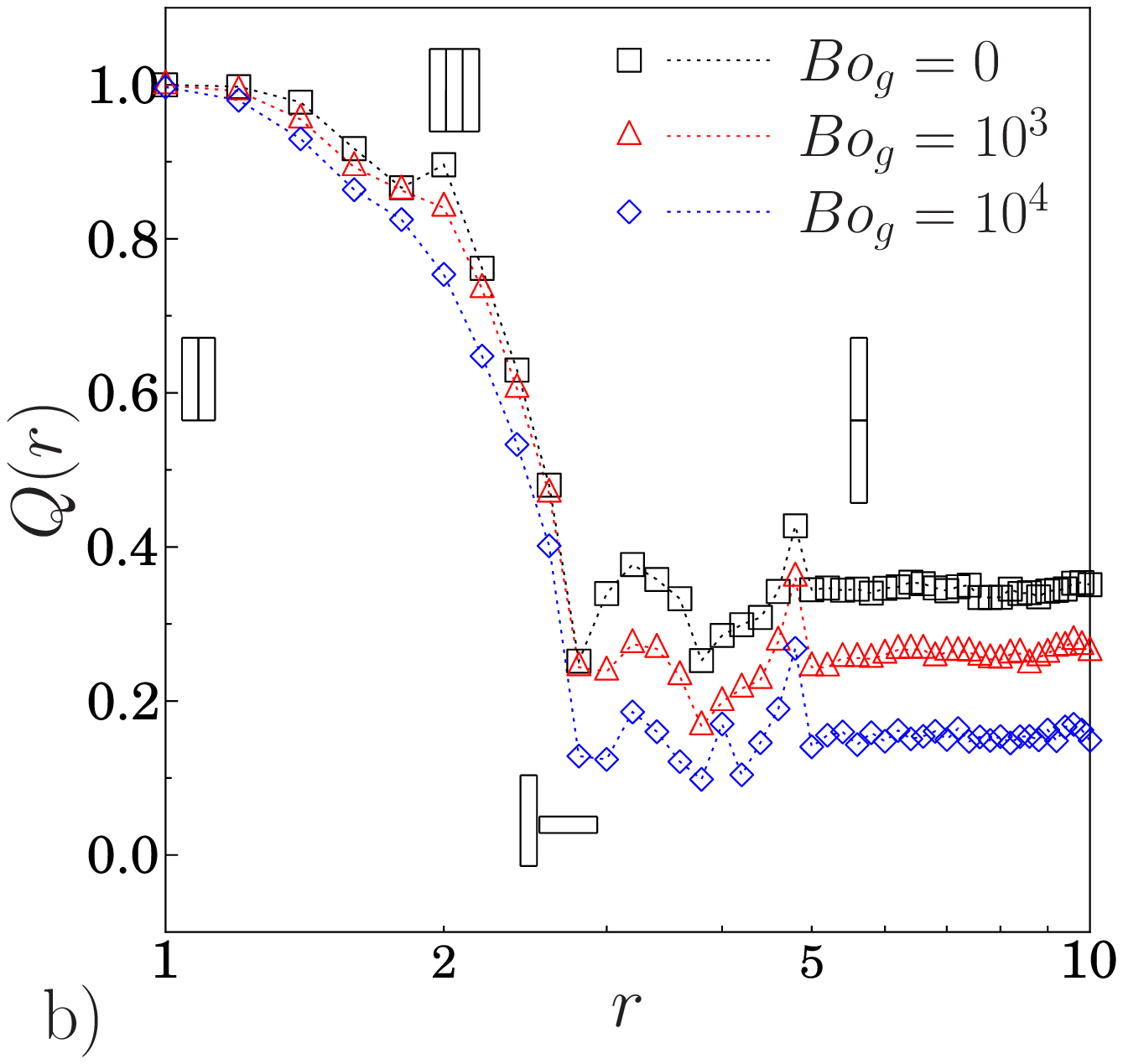}
   \caption{(Color online) Radial orientation distribution functions,
     $Q(r)$, as defined in Eq. \ref{eq_f} 
for rod deposits with two different aspect ratios, a) $d=2$ and b) $d=5$.
In each case, results for several Bond numbers are presented.  At specific
   maxima/minima the corresponding particle configurations are illustrated.}
\label{fig:Correlation_Q}
\end{center}
\end{figure}

\begin{figure}
\begin{center}
\includegraphics[width=6.3cm]{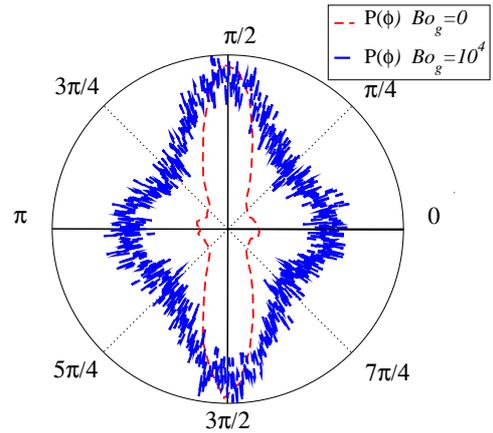}
\end{center}
\vspace{1em}
   \caption{(Color online) Polar distribution of the principal
direction (larger eigenvalue of the stress tensor for each particle $\sigma_{\alpha \beta}$), 
obtained for particles with $d=10$. For comparison, the data for $Bo_g=0$ and $Bo_g=10^4$  
is presented.}
\label{polar}
\end{figure}

The  packing morphology is also examined through a 
radial orientation function $Q(r)$, defined as
\begin{equation}
Q(r) = \langle  \cos{( 2(\theta_i -\theta_j))} \delta({\bf
    r}_{ij}-r) \rangle
\label{eq_f}
\end{equation}
where $\theta_i$  and  $\theta_j$ are the angular orientations of
particles $i$ and $j$, respectively. $Q(r)$ accounts for the mean value 
of the angular correlation between a given particle $i$ and a particle $j$ 
with their center of mass at a distance $r_{ij}$. 
 Note that, this distribution function 
provides useful quantitative information on the local morphology of the rod packings. 
Configurations where the two rods are perpendicular to each other
contribute $-1$ to $Q(r)$, while rods aligned along their long faces 
or along their short faces contribute with $1$ \cite{raulprl2009,raul10a}.

In Fig\ref{fig:Correlation_Q}a and Fig\ref{fig:Correlation_Q}a 
the numerical data for packing of cohesive particles 
with aspect ratio $d=2$ and $d=5$ are shown for comparison. 
Both figures show a series of maxima (parallel alignment) 
and minima (perpendicular alignment), which develop at several
distances.  This correlates with the high tendency of the particles to align 
in closely packed structures, in particular for low attractive force strengths 
and the limiting case of non-cohesive particles \cite{raulprl2009,raul10a}. 
 As the strength of the interaction force gets larger looser structures 
are formed and consequently the intensity of the maxima and minima decreases.
On the other hand, for very elongated particles  (see Fig.~\ref{fig:Correlation_Q}.b) 
$Q(r)$ clearly  indicates that the system does not show significant order.   
There is a maximum  at contact, which  simply indicates that at this shortest distance 
only perfectly aligned particles along their long faces can contribute. 
At larger distances a  deep plateau seems to indicate a small 
preference at these intermediate distances to observed parallel aligned particles.
Only a characteristic peak corresponding to the contact through the particle
\dkc{ length}  $r=d$ 
seems to remain even for very strong attractive strengths.    
This picture suggests the lack of significant order and presence of 
strong density inhomogeneities, in granular deposits of very cohesive 
particles. As aspect ratio gets higher, this effect is notably enhanced.  

Furthermore, we can correlate the microstructure with stress transmission
by studying the micro-mechanical properties of the granular deposits.
To this end, we introduce the stress tensor of a single particle $i$,
\begin{equation}
\sigma^i_{\alpha \beta} = \sum_{c=1}^{C_i} l_{i,\alpha}^c F_{i,\beta}^c,
\label{eq:stress}
\end{equation}
which is defined in terms of the total contact force $\vec{F}_i^c$
that particle $i$ experiences at contact $c$ and  
the branch vector $\vec{l}_i^c$  related to the contact $c$. The sum runs over
all the contacts $C_i$ of particle $i$, $\alpha,\beta$ are the
vectorial components. 

In Figure \ref{polar}, the polar distribution $P(\phi)$ of the
principal direction $\phi$ related to the larger eigenvalue of $\sigma_{\alpha \beta}$, 
obtained for particles with aspect ratio $d=10$ is illustrated.
For the non-cohesive case ($Bo_g=0$), Fig.\ \ref{polar} indicates 
that forces are preferentially transmitted in the vertical direction, displaying 
a high degree of alignment with the external gravity field.
Note, that the stress is dominated by the contribution parallel 
to gravity ($\sigma_{11}$) whose mean value (data not shown) 
is also much higher than the stress in the horizontal direction 
($\sigma_{22}$).
For very cohesive particles (see Fig\ref{polar}), however, 
the polar distribution of the principal direction is more uniform, 
denoting the establishment of a more spherical stress state.
Hence, in this case the stress is more isotropically transmitted while the 
alignment with the external gravity field diminishes. This effect correlates 
with the formation of very loose packings and the lack of significant order 
and the presence of strong density inhomogeneities.
\begin{figure}
 \begin{center}
\includegraphics[width=7.6cm]{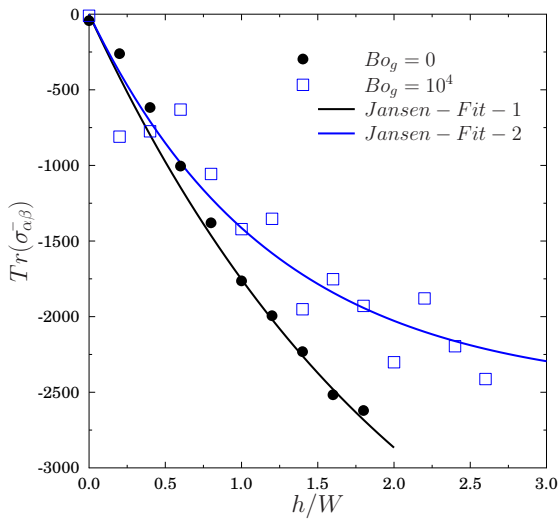}
   \caption{Profiles of the trace of the mean stress tensor, 
obtained for particles with $d=10$ and  two different bond numbers. The depth is normalized with the width $W$ of the system.
We also show the fitting to the equation $\sigma = \sigma_m (1-exp(-x/h_s))$,  
where we used $[\sigma_m=2500 N/m;h_s/W=1.2]$ and $[\sigma_m=4800 N/m;h_s/W=2.2]$ for case $1$ and $2$, respectively. 
} 
\label{fig:jansen}
\end{center}
\end{figure}

Finally, we show that the changes in microstructure induced both by particle 
 geometry and  the attractive force lead to significant modifications in the
 pressure (trace of the mean stress tensor)  profiles  as a function
of the silo depth $h$. The mean stress tensor, $\bar{\sigma}_{\alpha \beta}$, 
can be calculated for a given representative volume element (RVE) with area $A_{\rm RVE}$ resulting in  
\begin{equation}
\bar{\sigma}_{\alpha \beta} = \frac{1}{A_{\rm RVE}} \sum^N_{i=1}  w_v \sigma_{\alpha \beta}^i.
\label{eq:StressT}
\end{equation}
\noindent The sum runs over 
the representative volume element while
$w_v$ is an appropriate average weight. Although recently 
Professor Isaac  Goldhirsch and coworkers have developed a very accurate procedure 
for calculating $w_v$ and $\bar{\sigma}_{\alpha \beta}$ \cite{go1,go2}, 
for simplicity's sake we use particle-center averaging and choose the 
simplest weighting: $w_v=1$ if the center of the particle lies
inside the averaging area $A_{\rm RVE}$ and $w_v = 0$ otherwise~\cite{latzel00a,luding_madadi}. 

In Fig.~\ref{fig:jansen}, we display the trace of the mean stress tensor, defined following
Eq.~\ref{eq:StressT}, for elongated particles with $d=10$. For the 
calculation of the mean stress tensor we have used a representative volume 
element with a size equivalent to five particle lengths $A_{\rm RVE}=5d\times5d$.
The depth has been normalized with the width of the deposit $x=h/W$.
Moreover, for comparison, the numerical fit using a {\it Janssen-type} 
formula $\sigma = \sigma_m (1-exp(-x/h_s))$, are also
shown. 
Here, the magnitude of $\sigma_m$ represents the saturation stress and $h_s$ indicates 
the characteristic value of depth at which the pressure in the deposit
stabilizes.  
As we have mentioned earlier, non-cohesive elongated particles 
transmit stress preferentially parallel to gravity. As a result, the weak
transmission to the the lateral walls weakens pressure saturation (Fig.~\ref{fig:jansen}), 
which is reflected by the large saturation depth $h_s$ and stress $\sigma_m$.
The latter is consequence of the horizontal alignment of the flat faces of the rods, which
induces an anisotropic stress transmission, from top to bottom.
However, the scenario changes drastically for very cohesive particles. 
As we also pointed out earlier, when the attractive force is increased, particle orientations 
deviate from the horizontal and a larger disorder in the particle orientation
distribution shows up. As a result, the spherical component of the local stress 
is notably enhanced with respect to the deviatoric part, which is related 
to the gravity direction.  Hence, for very cohesive particles $B_o=10^4$ of $d=10$  we found 
notably smaller values of saturation depth $h_s$ and stress $\sigma_m$. In
this respect, introducing an attractive force has a very similar effect as reducing the particle elongation.    

\section{Conclusion}
We  have shown that introducing an attractive force in deposits 
of elongated grains has a profound effect on the deposit morphology and 
its stress profiles. In deposits of non-cohesive particles the topology is 
dominated by the formation  of ordered structures of aligned rods. Elongated particles 
tend to align horizontally and  the stress is mainly transmitted from top to
bottom, revealing an asymmetric distribution of the local stress.  Lateral
force transmission becomes less favored compared to vertical transfer,
thus hindering pressure saturation 
with depth.  For deposits of cohesive particles, the
preferred horizontal orientation  is less pronounced with increasing
 cohesion.  Very elongated particles with strong attractive forces form extremely 
loose structures, characterized by orientation distributions, which  
 tend  to a uniform behavior when increasing the Bond number. As a result of these 
changes, the pressure distribution in the deposits changes qualitatively. 
The spherical component of the local stress is notably enhanced with respect to 
the deviatoric part. Hence, the lateral stress transmission is promoted by the 
enhanced disorder and it leads to a faster pressure saturation with depth. 

\section*{Acknowledgment}
The Spanish MICINN Project FIS2008-06034-C02-02 has supported this work. This work started while the author RCH was visiting the IfB at the Eidgen\"ossische Technische Hochschule of Z\"urich. Its financial support and hospitality are gratefully acknowledged. TK acknowledge the University of Girona (Spain) for financial support.

\end{document}